\documentclass[twocolumn,letter]{jpsj3}
\usepackage[dvipdfmx]{graphicx, color}
\usepackage{bm}
\usepackage{cite}
\usepackage{color}
\usepackage{amssymb}
\usepackage{amsmath}
\usepackage{revsymb}
\usepackage{ulem}

\def\<{\langle}
\def\>{\rangle}
\def\({\left(}
\def\){\right)}
\def\[{\left[}
\def\]{\right]}
\def\up{\uparrow}
\def\dn{\downarrow}

\def\+{\dagger}

\title{Excitonic Order and Superconductivity in the Two-Orbital Hubbard Model: Variational Cluster Approach}

\author{
Ryo Fujiuchi$^1$\thanks{r.fujiuchi\_104@chiba-u.jp},
Koudai Sugimoto$^2$, and Yukinori Ohta$^1$}
\inst{
$^1$Department of Physics, Chiba University, Chiba 263-8522, Japan \\
$^2$Center for Frontier Science, Chiba University, Chiba 263-8522, Japan \\} %\\

\date{\today}

\abst{
Using the variational cluster approach based on the self-energy functional theory, we study the possible 
occurrence of excitonic order and superconductivity in the two-orbital Hubbard model with intra- and 
inter-orbital Coulomb interactions.  It is known that an antiferromagnetic Mott insulator state appears in 
the regime of strong intra-orbital interaction, a band insulator state appears in the regime of strong 
inter-orbital interaction, and an excitonic insulator state appears between them.  
In addition to these states, we find that the $s^{\pm}$-wave superconducting state appears in 
the small-correlation regime, and the $d_{x^2 - y^2}$-wave superconducting state appears on 
the boundary of the antiferromagnetic Mott insulator state.  
We calculate the single-particle spectral function of the model and compare the band gap formation 
due to the superconducting and excitonic orders.  
}

\begin{document}
\maketitle

% ----------Introduction----------

Multi-orbital superconductivity arising from purely electronic mechanisms has thus far been discussed 
primarily in the context of heavy-fermion systems and transition-metal compounds.  In the former, 
strongly correlated $f$-electrons hybridized with itinerant conduction electrons can give rise 
to superconductivity, where the periodic Anderson model has often been used for theoretical 
studies \cite{Rice1985PRL, Ikeda2006JPSJ}.  
In the latter, the orbital degrees of freedom of transition-metal ions play an important role in the 
superconductivity.  In iron-based superconductors, for example, antiferrmagnetic and orbital fluctuations 
in the presence of multiple Fermi surfaces made of $3d$ orbitals have been argued to cause 
superconductivity with either $s^{\pm}$ or $s^{++}$ pairing symmetry 
\cite{Hirschfeld2011RPP, Hosono2015PC, Vorontsov2009PRB, Fernandes2010PRB, Maiti2010PRB}.  
Further, in Sr$_2$RuO$_4$, multiple Fermi surfaces consisting of the $t_{2g}$ orbitals have been 
predicted to cause spin-triplet pairing superconductivity \cite{Maeno2012JPSJ}.  
In theoretical studies of these systems, multi-orbital Hubbard models with on-site interactions 
have typically been used \cite{Koga2002PRB, Han2004PRB, Sakai2004PRB, Kubo2007PRB, 
Sano2009JPSJ, Maier2011PRB, Koga2015PRB}.  

Recently, a possibly different type of multi-orbital superconductivity, which appears adjacent to 
the excitonic order (or excitonic-insulator state) \cite{Jerome1967PR, Halperin1968RMP}, was 
reported to occur in transition-metal chalcogenides.  A candidate material $1T$-TiSe$_2$ 
\cite{DiSalvo1976PRB, Pillo2000PRB, Kidd2002PRL, Cercellier2007PRL, Kogar2017Science} 
shows superconductivity either when pressure is applied \cite{Kusmartseva2009PRL} 
or when Cu atoms are intercalated \cite{Morosan2006NP, Qian2007PRL, Li2007PRL, Zhao2007PRL}.  
The Fermi surfaces come from the Ti $3d$ and Se $4p$ orbitals, and nesting of these multiple 
Fermi surfaces leads to excitonic order \cite{Koley2014PRB, Kaneko2017arXiv}.  
Another candidate material, Ta$_2$NiSe$_5$~\cite{Wakisaka2009PRL, Seki2014PRB, Kim2016ACSN, 
Lu2017NC, Larkin2017PRB, Mor2017PRL, Seo2017arXiv}, also shows superconductivity under applied 
pressure \cite{Matsubayashi2018}.  This material is a semiconductor with a small direct band gap 
between the Ni $3d$ valence band and Ta $5d$ conduction band at the $\Gamma$ point 
of the Brillouin zone \cite{Kaneko2013PRB}.

Not much is known, however, about the competition between superconductivity and excitonic order.  
Although such studies are of general importance in the field of condensed matter physics and 
should be developed using multi-band models such as multi-orbital Hubbard models, only a limited 
number of studies have been made to date of Hubbard-like lattice models, which include attractive 
on-site \cite{Sarasua2002PRB} and inter-site \cite{Vanhala2015PRB} interactions for specific 
purposes.  

In this paper, we present a study of the excitonic order and superconductivity in the two-orbital 
Hubbard model using the variational cluster approach (VCA) \cite{Potthoff2003PRL} based on the 
self-energy functional theory \cite{Potthoff2003EPJB}.  Thus, we can treat the ordered phases 
of the model that occur because of spontaneous symmetry breaking, such as antiferomagnetic order 
\cite{Dahnken2004PRB}, excitonic order \cite{Seki2014PRB, Kaneko2012PRB, Kaneko2014PRB, 
Kaneko2015PRB}, and superconducing order \cite{Senechal2005PRL, Aichhorn2006PRB, 
Sahebsara2006PRL, Nevidomskyy2008PRB, Kaneko2014JPSJ, Masuda2015PRB}, on an equal footing.  
We calculate the phase diagram of the model in the parameter space of the intra-orbital and 
inter-orbital Coulomb interactions and show that the antiferromagnetic Mott insulator (AFMI) 
state appears in the regime of strong intra-orbital interaction, the band insulator (BI) state 
appears in the regime of strong inter-orbital interaction, and, between the two, excitonic order 
appears, as was shown in Ref.~\cite{Kaneko2012PRB}.  
Moreover, we find that, in addition to these states, the spin-singlet $s^{\pm}$ superconducting 
($s^{\pm}$SC) state appears in the small-correlation regime and the spin-singlet $d_{x^2 -y^2}$ 
superconducting ($d$SC) state appears on the boundary of the AFMI phase.  
The competition between superconductivity and excitonic order is then discussed.  
We also calculate the single-particle spectral functions of the model and compare the band 
gap formation due to these superconducting and excitonic orders.  

% ----------Model and Method----------

% \section{Model and method}
% \label{sec:model_and_method}
% \subsection{Two-orbital Hubbard model}

\begin{figure}[tb]
\begin{center}
\includegraphics[width=8.5cm]{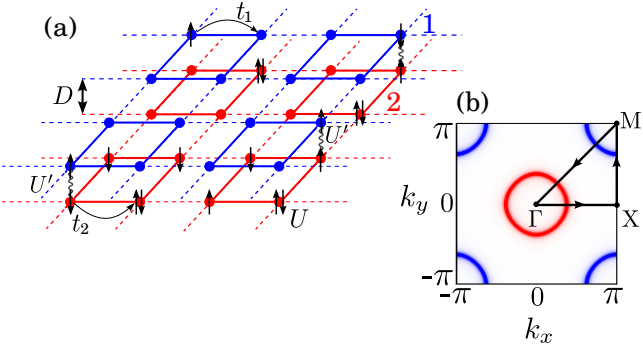}
\end{center}
\caption{(Color online) 
(a) Schematic representation of the two-orbital Hubbard model defined on a two-dimensional 
square lattice, where we assume the presence of intra- and inter-orbital Coulomb interactions 
($U$ and $U'$, respectively).  There are no hopping integrals between the orbitals $\mu = 1$ (blue) 
and 2 (red).  
(b) Fermi surfaces of the model in the noninteracting limit.  A hole pocket and an electron 
pocket are located at the $\Gamma = (0,0)$ and $M=(\pi, \pi)$ points of the Brillouin zone, 
respectively.  
}\label{fig:model}
\end{figure}

We consider the two-orbital Hubbard model defined on a two-dimensional square lattice of 
lattice constant $a=1$ [see Fig.~\ref{fig:model}(a)].  We assume that the energies of the two orbitals 
are separated by $D$ and that there are no hopping integrals between the two orbitals.  
Repulsive Coulomb interactions occur between electrons on the two orbitals.  
The Hamiltonian is written as $H = H_0 + H_U + H_{U'}$ with
\begin{align}
&H_0 = \sum_{\< i, j \>, \mu, \sigma} t_{\mu} c_{i, \mu, \sigma}^\+ c_{j, \mu, \sigma} \notag \\
&~~~~~~~+ \frac{D}{2} \sum_{i, \sigma} \( c_{i, 1, \sigma}^\+ c_{i, 1, \sigma} - c_{i, 2, \sigma}^\+ c_{i, 2, \sigma} \),
\\
&H_U = U \sum_{i, \mu} \( n_{i, \mu, \up} - \frac{1}{2} \) \( n_{i, \mu, \dn} - \frac{1}{2} \),
\\
&H_{U'} = U' \sum_{i, \sigma, \sigma'}  \( n_{i, 1, \sigma} - \frac{1}{2} \) \( n_{i, 2, \sigma'} - \frac{1}{2} \), 
\label{eq:H_U'}
\end{align}
where $c_{i, \mu, \sigma}$ ($c_{i, \mu, \sigma}^\+$) is the annihilation (creation) operator of 
an electron with spin $\sigma$ at orbital $\mu \; (=1,2)$ and site $i$.  
Further, $t_\mu$ is the hopping integral between the orbitals $\mu$ and $\<i,j\>$ represents the 
nearest-neighbor pair of sites $i$ and $j$.  We take into account the intra-orbital ($U$) and 
inter-orbital ($U'$) Coulomb interactions.  Other interactions, such as Hund's rule coupling 
and the pair hopping interaction, are not taken into account.  We set $t_1 = t_2 = t = 1$ as 
the unit of energy and assume $D/t=6$, so that the system is semimetallic in the noninteracting 
limit (or at $U=U'=0$).  The electron filling is assumed to be 
$\sum_{\mu, \sigma} \langle n_{i, \mu, \sigma} \rangle = 2$, 
so that the system has electron-hole symmetry.  The Fermi surfaces without the interactions 
are shown in Fig.~\ref{fig:model}(b).  There is perfect nesting between the electron and hole 
Fermi surfaces.  

% \subsection{Weiss fields}\label{sec:ex_cond}

To consider the spontaneous symmetry breaking of the system within the framework of the VCA, 
we add the Weiss field $H' = h \sum_{i} O_i$ to the Hamiltonian, where $h$ is the strength of the 
Weiss field, and $O_i$ is the corresponding single-particle operator defined in real space.  
We define the order parameter as $\Delta = \frac{1}{L} \sum_{i} \langle O_i \rangle$, where $L$ is 
the number of sites in the system.  We consider the AFMI, excitonic charge-density-wave (ECDW), 
and spin-singlet superconducting orders, for which we define the Weiss fields as follows: 
\begin{equation}
 H'_{\rm AFM} = h_{\rm AFM} \sum_{i, \mu, \sigma} \sigma e^{i \bm{Q} \cdot \bm{r}_i} c^{\dagger}_{i, \mu, \sigma} c_{i, \mu, \sigma}
\end{equation}
for the AFMI order, 
\begin{equation}
 H'_{\rm ECDW} = h_{\rm ECDW} \sum_{i, \sigma} e^{i \bm{Q} \cdot \bm{r}_i} c^{\dagger}_{i, 1, \sigma} c_{i, 2, \sigma} + {\rm H.c.}
\end{equation}
for the ECDW order,
\begin{equation}
 H'_{s^{\pm}} = h_{s^{\pm}} \sum_{\mu} \sum_{\langle i, j \rangle} c_{i, \mu, \uparrow} c_{j, \mu, \downarrow} + \rm{H.c.} 
\end{equation}
for the $s^{\pm}$SC order, and
\begin{align}
&H'_{d_{x^2 - y^2}}
	= h_{d_{x^2 - y^2}} \sum_{\mu} \( \sum_{\langle i, j \rangle_x}
		c_{i, \mu, \uparrow} c_{j, \mu, \downarrow}
		- \sum_{\langle i, j \rangle_y}
		c_{i, \mu, \uparrow} c_{j, \mu, \downarrow} \)
\nonumber\\
&~~~~~~~~~~~~~~+ \rm{H.c.}
\end{align}
for the $d$SC order, where $\langle i, j \rangle_\alpha$ denotes the nearest-neighbor pair 
along the $\alpha$ $(=x, y)$ direction.  
We assume the ordering vector $\bm{Q} = (\pi, \pi)$ for the AFMI and ECDW orders.  
We consider only the ECDW order because the energies of the ECDW and excitonic 
spin-density-wave orders are exactly degenerate in the absence of Hund's rule 
coupling \cite{Kaneko2014PRB, Buker1981PRB}.  

We also consider the inter-orbital superconductivity, where Cooper pairs are formed between 
orbitals 1 and 2 \cite{Kubo2007PRB, Sano2009JPSJ, Vanhala2015PRB}.  
The Weiss field is written as
\begin{equation}
H' = h \sum_{i} e^{i \bm{Q} \cdot \bm{r}_i} \Big(c_{i, 1, \uparrow} c_{i, 2, \downarrow} - c_{i,1,\downarrow} c_{i,2,\uparrow}\Big)+ {\rm H.c.},  
\end{equation}
where $\bm{Q} = (\pi, \pi)$ is the momentum of the Cooper pair.  
Although the ECDW fluctuation enhances the effective pairing interaction of the inter-orbital Cooper pairs, 
we find that this order is not stable in our model for the following reason.  
To form excitonic pairs, there should be at least one electron Fermi surface and one hole Fermi surface.  
On the other hand, to form superconducting Cooper pairs, two electron Fermi surfaces (consisting of the 
up and down spins) or two hole Fermi surfaces are required.  In our model, there are two spin-degenerate 
electron Fermi surfaces consisting of orbital 1 and two spin-degenerate hole Fermi surfaces consisting 
of orbital 2.  This means that Cooper pairs can be formed only between electrons (or holes) on the same 
orbital.  We confirmed numerically that interorbital Cooper pairing is not stable in the present two-orbital 
Hubbard model.  

In addition to these states, we also investigated other types of superconductivity, i.e., $s^{++}$ 
superconductivity (on-site pairing) and spin-triplet superconductivity.  By comparing their grand potentials 
(as discussed below), we found that these types of superconductivity are not stable in the entire parameter 
space examined.  

We employ the VCA \cite{Potthoff2003PRL}, adopting a reference system consisting of $L=2 \times 2$ 
square clusters [see Fig.~\ref{fig:model}(a)].  The grand potential of the original system may then be 
written as 
\begin{equation}
\Omega
	= \Omega' + \frac{1}{\beta} \sum_n \mathrm{Tr} \ln \[ \mathcal{G}_0^{-1} (i \varepsilon_n) 
- \Sigma \]^{-1}
	- \frac{1}{\beta} \sum_n \mathrm{Tr} \ln \mathcal{G} (i \varepsilon_n)
\end{equation}
where $\Omega'$ is the grand potential of the reference system, and $\varepsilon_n$ is the fermionic 
Matsubara frequency.  $\mathcal{G}$ and $\Sigma$ are the temperature Green's function and self-energy 
of the reference system, respectively, which are calculated by exact diagonalization of small clusters, 
and $\mathcal{G}_0$ is the noninteracting temperature Green's function of the original system.  

The strength of the Weiss field is determined by minimizing the grand potential \cite{Dahnken2004PRB}, i.e., 
\begin{equation}
 \frac{\partial \Omega}{\partial h} = 0.
 \label{eq:stationary_point}
\end{equation}
When $\Omega$ has a stationary point at a finite Weiss field $h$, the system goes into the 
spontaneous symmetry breaking state.  We consider the Weiss fields defined above and compare 
the stationary values of the grand potential to determine the ground-state phase diagram of the 
system.  For simplicity, the possible coexistence of different orders is not taken into account.  

\begin{figure}[tbp]
\begin{center}
\includegraphics[width=8.5cm]{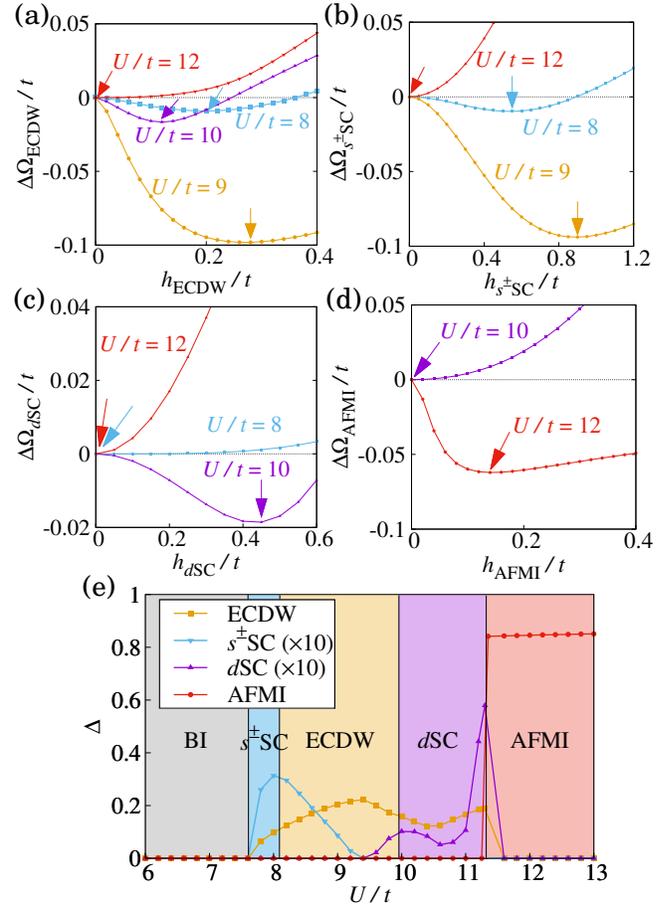}
\end{center}
\caption{(Color online) 
Calculated grand potential as a function of the Weiss fields for 
(a) ECDW, (b) $s^{\pm}$SC, (c) $d$SC, and (d) AFMI orders.  The arrows indicate the stationary points.
(e) Calculated order parameters as a function of $U/t$.  
We assume $U'/t=5$ and $D/t=6$ in all panels.  
}\label{fig:op}
\end{figure}

Now, let us discuss the results of the calculations.  
First, it is intuitively clear that the BI phase is stable in the large-$U'$ regime because two electrons 
favorably occupy the same orbital at a site, and that the AFMI phase is stable in the large-$U$ regime 
because two electrons favorably occupy different orbitals at a site (aligning the spin directions in 
parallel in the presence of Hund's rule coupling).  It was also shown that, in the intermediate 
regime between these two phases, the excitonic phase becomes lower in energy than the paramagnetic 
metallic phase \cite{Kaneko2012PRB}.  However, the possible stability of the superconducting phase 
in this regime has not been fully studied to date.  Then, we perform detailed VCA calculations to 
seek superconducting phases in this regime.  

In Figs.~\ref{fig:op}(a), \ref{fig:op}(b), \ref{fig:op}(c), and \ref{fig:op}(d), we show the calculated 
differences in the grand potentials $\Delta \Omega = \Omega(h) - \Omega(h=0)$ as a function of 
the Weiss fields for the ECDW, $s^{\pm}$SC, $d$SC, and AFMI orders, respectively.  We find 
that the grand potential has a stationary point for all the orders.  
In Fig.~\ref{fig:op}(e), we show the order parameters calculated for the $s^{\pm}$SC, $d$SC, 
ECDW, and AFMI phases as a function of $U$, where we assume $U' / t = 5$.  
All the phases (BI, $s^{\pm}$SC, $d$SC, ECDW, and AFMI) appear as $U$ is varied.  The phase 
boundaries are determined by comparing the grand potentials of these phases.  We note that the 
transitions between these ordered phases are of the first order because the system with different 
$U$ values has long-range order with different broken symmetries. 
We also note that the number of electrons in each orbital within the small cluster used changes 
by one just at the phase boundary between the $d$SC and AFMI phases. This means that the location 
of this phase boundary in the calculated phase diagram is subject to the finite-size effect of the 
small cluster used.  

\begin{figure}[tbp]
\begin{center}
\includegraphics[width=7.5cm]{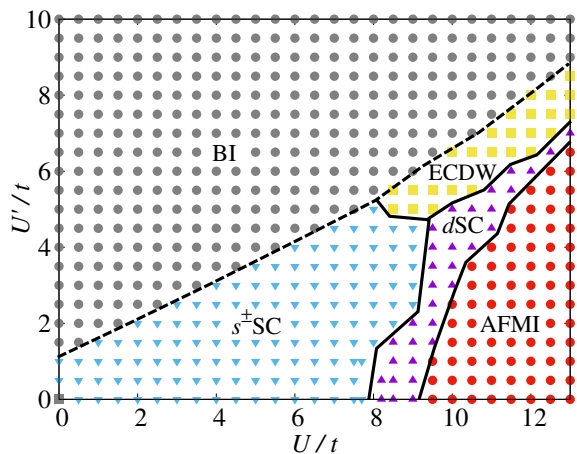}
\end{center}
\caption{(Color online) 
Calculated ground-state phase diagram of the two-orbital Hubbard model at $D / t = 6$.  
The $s^{\pm}$SC (ECDW) phase appears between the AFMI and BI phases in the weaker (stronger) 
interaction regime.  The $d$SC phase also appears on the boundary of the AFMI phase.  
The solid and dotted lines indicate the first-order (discontinuous) and second-order (continuous) 
phase transitions, respectively.  
}\label{fig:phase}
\end{figure}

In Fig.~\ref{fig:phase}, we show the calculated ground-state phase diagram of our model in the 
$(U,U')$ parameter space.  We again confirm that the system is in the AFMI phase when $U \gg U'$, 
whereas it is in the BI phase when $U \ll U'$.  The ECDW phase appears between these 
two phases, just as in a previous study \cite{Kaneko2012PRB}.  
We then find that the $s^{\pm}$SC phase becomes lower in energy than the ECDW phase in the 
weak-correlation regime \cite{note1}.  In addition, we also find that the $d$SC phase emerges 
on the boundary of the AFMI phase.  This means that the reduction of the grand potential due 
to ECDW formation is small (large) in the weak-correlation (strong-correlation) regime, so the 
gain in energy due to superconducting pair formation becomes larger than the gain due to ECDW 
formation.  

\begin{figure*}[tbh]
\begin{center}
\includegraphics[width=15.5cm]{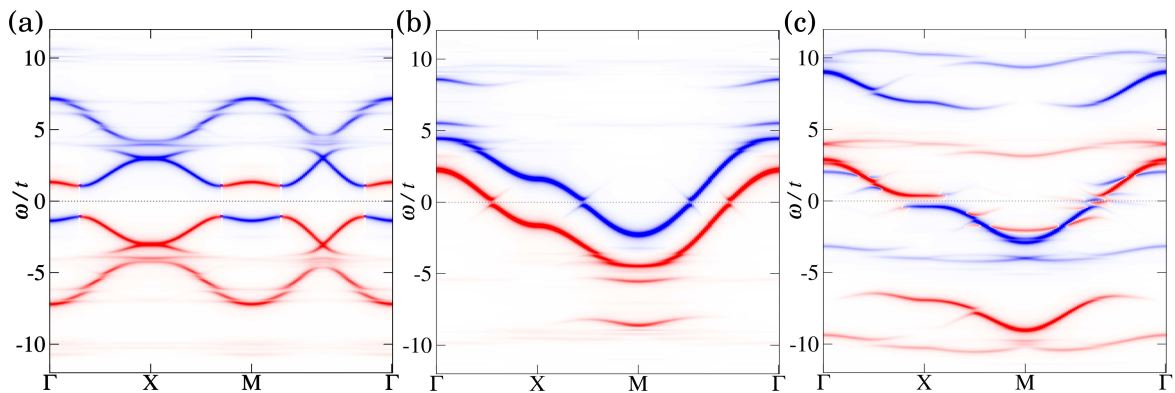}
\end{center}
\caption{(Color online) 
Calculated single-particle spectra of our model
(a) in the ECDW phase at $U/t=9$ and $U'/t = 5$,
(b) in the $s^\pm$SC phase at $U/t=6$ and $U'/t = 2$, and
(c) in the $d$SC phase at $U/t=9$ and $U'/t = 1$.
The orbital 1 (2) dominant spectra are depicted in blue (red).  
}\label{fig:spectra}
\end{figure*}

The emergence of these superconducting phases may be attributed to the antiferromagnetic 
spin fluctuations as follows.  For large $U$ and small $U'$, the lower Hubbard band of orbital 1 
becomes lower in energy than the upper Hubbard band of orbital 2, and the system becomes an 
AFMI.  The inter-site interaction $U'$ pushes the band of orbital 1 up (i.e., causes the Hartree 
shift), so the band of orbital 1 supplies electrons to the band of orbital 2.  This situation is the 
same as that in doped Mott insulators in the single-band Hubbard model \cite{Senechal2005PRL}, 
where the AFMI order is suppressed, and the superconducting order driven by antiferromagnetic 
spin fluctuations emerges \cite{ Bulut1992PRB, Kuroki2002PRB}.  We also note that the switching 
behavior between the $s^{\pm}$SC and $d$SC states that we found is consistent with a previous 
study of the bilayer Hubbard model~\cite{Maier2011PRB}.  
We will clarify the origin of the $s^\pm$SC order in the weak-correlation regime in future 
using the perturbative weak-coupling approach, in which the instability toward the superconducting 
state is examined.  

Finally, let us discuss the band-gap opening characteristic of the superconducting and excitonic 
orders, which manifests itself in the calculated single-particle spectral function defined as 
$A(\bm{k}, \omega) = -\frac{1}{\pi} {\rm Im} \, G(\bm{k}, \omega)$, where $G$ is the retarded 
Green's function obtained from the cluster perturbation theory (CPT) in the superconducting 
and excitonic phases.  
The detailed numerical techniques for calculating the CPT Green's functions in the ordered 
states are found in, e.g., Refs.~\cite{Kaneko2012PRB, Kaneko2014JPSJ}.  
The calculated results are shown in Fig.~\ref{fig:spectra}.  
In the ECDW phase, the conduction and valence bands are folded, and the electron and hole 
Fermi surfaces overlap, at which the excitonic band gap opens isotropically in the entire 
${\bm k}$ space.  In the superconducting phases with intersite Cooper pairing, on the other hand, 
the order parameter of the $s^{\pm}$SC phase depends on the wave vector and has nodes.  
The nodes are located between the electron and hole Fermi surfaces, so a finite band-gap (or 
Bardeen--Cooper--Schrieffer gap) opens on the entire Fermi surface.  The $d$SC phase also 
has a $\bm{k}$-dependent order parameter, where the nodes appear along the M-$\Gamma$ 
line of the Brillouin zone; consequently, the band gap closes at the nodes as seen in 
Fig.~\ref{fig:spectra}(c).  The size of these band gaps reflects the strength of the order parameters.  

In summary, we studied the possible occurrence of superconductivity in the two-orbital Hubbard model 
with intra- and inter-orbital Coulomb interactions using the VCA based on the self-energy 
functional theory.  
We first confirmed that the AFMI state appears in the regime with strong intra-orbital interaction, 
the BI state appears in the regime with strong inter-orbital interaction, and the excitonic insulator 
state appears between them.  
We then carefully examined the possible occurrence of superconductivity in the intermediate 
interaction regime and found that the $s^{\pm}$SC state is in fact lower in energy compared to 
the excitonic insulator state, particularly in the weak-correlation regime.  
In addition, we found that the $d$SC state appears on the boundary of the AFMI state.  
The appearance of these superconducting states may be attributed to antiferromagnetic spin 
fluctuations.  We also calculated the single-particle spectral function of the model and discussed 
the band gap formations due to the superconducting and excitonic orders.  

\medskip
\begin{acknowledgment}
% people
We thank T. Kaneko, K. Masuda, S. Miyakoshi, H. Nishida, and T. Yamaguchi for enlightening discussions.  
% grant
This work was supported in part by a Grant-in-Aid for Scientific Research (No.~17K05530) from JSPS of Japan.  
\end{acknowledgment}

%\end{references}


\begin{thebibliography}{99}
\bibitem{note1} The phase diagram indicates that the $s^\pm$SC state appears even at $U'=0$.  However, because 
our model tends toward two independent doped single-band Hubbard models at the $U'\rightarrow 0$ limit, we do not 
expect $s$-wave superconductivity to occur there.  Note that the order parameter for the $s^\pm$SC state 
decreases to 0 with decreasing $U'$, but owing to weakness of the VCA when a small solver cluster is used, 
the order parameter does not vanish even at $U'=0$.
\end{thebibliography}
\end{document}